\documentclass{elsarticle}

\usepackage{graphicx}
\title{Dark Matter Spin-Dependent Limits for WIMP Interactions 
on $^{19}$F by PICASSO}

\author[mtl]{S. Archambault}
\author[mtl]{F. Aubin\fnref{mcgill}}
\author[mtl]{M. Auger \fnref{bern}} 
\author[iusb]{ E. Behnke} 
\author[ab]{ B. Beltran} 
\author[qu]{ K. Clark\fnref{case} } 
\author[qu]{ X. Dai \fnref{criver}} 
\author[qu]{ A. Davour} 
\author[lu]{ J. Farine} 
\author[mtl]{ R. Faust } 
\author[mtl]{  M.-H. Genest \fnref{muc}} 
\author[mtl]{ G. Giroux \fnref{bern}} 
\author[mtl]{ R. Gornea \fnref{bern}} 
\author[ab]{ C. Krauss} 
\author[mtl]{  S. Kumaratunga} 
\author[snolab]{  I. Lawson } 
\author[mtl]{ C. Leroy} 
\author[mtl]{ L. Lessard} 
\author[qu]{ C. Levy} 
\author[iusb]{ I. Levine} 
\author[ab]{ R. MacDonald} 
\author[mtl]{ J.-P. Martin } 
\author[lu]{ P. Nadeau} 
\author[qu]{ A. Noble} 
\author[mtl]{  M.-C. Piro} 
\author[prague]{ S. Pospisil} 
\author[iusb]{ T. Shepherd} 
\author[mtl]{ N. Starinski} 
\author[prague]{ I. Stekl} 
\author[qu]{  C. Storey} 
\author[lu]{ U. Wichoski} 
\author[mtl]{ V. Zacek\corref{cor} }
\address[mtl]{D\'epartement de Physique, Universit\'e de Montr\'eal, Montr\'eal, H3C 3J7, Canada}
\address[qu]{Department of Physics, Queen's University, Kingston, K7L 3NG, Canada}
\address[ab]{Department of Physics, University of Alberta, Edmonton, T6G 2G7, Canada}
\address[lu]{Department of Physics, Laurentian University, Sudbury, P3E 2C6, Canada}
\address[iusb]{Department of Physics\&Astronomy, Indiana University South Bend, South Bend, IN 46634, USA}
\address[prague]{Institute of Experimental and Applied Physics, Czech Technical University in Prague, Prague, Cz-12800, Czech Republic}
\address[snolab]{SNOLAB, Sudbury, P3E 2C6}
\fntext[mcgill]{Present address: Department of Physics, McGill University, Montr\'eal, H3A 2T8, Canada}
\fntext[bern]{Present address: Laboratorium f\"ur Hochenergiephysik, Universit\"at Bern, CH-3012 Bern, Switzerland }
\fntext[case]{Present address: Department of Physics, Case Western Reserve University, Cleveland OH 44106-7079, USA}
\fntext[criver]{Present address: AECL Chalk River Laboratories, Chalk River ON K0J 1J0, Canada}
\fntext[muc]{Fakult\"at f\"ur Physik, Ludwig-Maximilians-Universit\"at, D-85748 Garching, Germany}
\cortext[cor]{Corresponding author: e-mail: zacekv@lps.umontreal.ca}

\begin{document}

\begin{abstract}
The PICASSO experiment at SNOLAB reports new results for spin-dependent WIMP interactions on $^{19}$F using the
superheated droplet technique. A new generation of detectors and new features which enable background discrimination 
via the rejection of non-particle induced events are described. 
First results are presented for a subset of two detectors with target masses of $^{19}$F of 65 g and 69 g respectively and a total exposure
of 13.75 $\pm$ 0.48 kgd. No dark matter signal was found and for WIMP masses around 24 GeV/c$^2$ new limits have been obtained on the spin-dependent cross section on $^{19}$F of $\sigma_F$  = 13.9 pb (90\% C.L.) which can be converted into cross section limits on protons and neutrons of $\sigma_p$ = 0.16 pb and $\sigma_n$ = 2.60 pb respectively (90\% C.L). The obtained limits on protons restrict recent interpretations of the DAMA/LIBRA annual modulations in terms of spin-dependent interactions.
\end{abstract}

\maketitle
\section{Introduction}\label{sec:intro}

The PICASSO experiment (Project In Canada to Search for Super-Symmetric Objects) searches for cold dark matter through the 
direct detection of neutralinos via their {\it spin-dependent} (SD) 
interactions with $^{19}$F nuclei \cite{ellis91,divari,bednyako}. 
The use of the light target nucleus $^{19}$F, together with a low detection threshold of 2 keV for recoil nuclei makes PICASSO particularly sensitive to low-mass WIMPs. This is important as few experiments are sensitive below 15 GeVc$^{-2}$ and it is precisely in this mass range where recent interpretations of the 
DAMA/LIBRA annual modulations in terms of SD interactions remain partially unchecked \cite{bernabei,savage}.\\   
    The elastic spin-dependent (SD) cross section for WIMP scattering at tree level and zero 
momentum transfer has the form \cite{levin96,engel91}
\begin{equation}\label{eq:1}
\sigma_{SD} = \frac{32}{\pi} G_F^2\left(\frac{M_WM_A}{M_W+M_A}\right)^2(a_p<S_p> +\, a_n <S_N>)^2 \frac{J+1}{J}
\end{equation}
where G$_F$ is the Fermi constant, M$_W$ and M$_A$ are the masses of the WIMP and the 
target nucleus, a$_{p,n}$ are the effective proton (neutron) coupling strengths, $<S_{p,n}>$ are the expectation values 
of the proton (neutron) spins in the target nucleus and J is the nuclear spin. \\
   With the exception of neutralino scattering on free protons, $^{19}$F is one of the most favourable target nuclei for the direct detection of SD interactions \cite{ellis91,beadnyakov97}.
As measurements and shell model calculations of nuclear magnetic moments show, the spin S$_{1/2}$ of the $^{19}$F nucleus is 
carried almost exclusively by its unpaired proton (e.g.  $<S_p>$ = 0.441 and $<S_n>$ = -0.109 in \cite{pachenko89}), 
which enhances the spin-dependent cross section in $^{19}$F by nearly an order of magnitude compared to other frequently 
used detector materials. Fluorine is also 
complementary to other SD target isotopes like $^{23}$Na, $^{127}$I and $^{129,131}$Xe in terms of the WIMP-proton ($a_p$) 
and WIMP- neutron ($a_n$) couplings. This is due to the negative sign in the ratio of the expectation values of the proton and  neutron spins which leads to a different range of sensitivities in the $a_p$--$a_n$ plane \cite{tovey,giulianiPRL04,giulianiPLB04}.\\
	The goal of the PICASSO project at SNOLAB is to exploit the favourable properties of $^{19}$F 
by using C$_4$F$_{10}$ as target material and by employing the superheated droplet detection technique, which is based on the operation principle of the classic bubble chamber \cite{glaser52,NC94}. Other dark matter experiments based on similar 
techniques are SIMPLE, using droplets of C$_2$ClF$_5$ and CF$_3$I \cite{girard05,giuliani07} and COUPP, 
which operates a bubble chamber filled with CF$_3$I \cite{behnke08}.
	Previous physics results of our studies have been published in \cite{hamel97,boukhira00}. In the meantime 
progress has been made in several areas and data are presently being taken with 32 detector modules with an 
increased active mass of $^{19}$F of about 65 g per module, with lower intrinsic background and an increased droplet size (section \ref{sec:det}). New discrimination variables, derived from the signal wave forms have been discovered, which allow an efficient suppression of non-particle induced backgrounds (section \ref{sec:disc}). 
In this paper we report the first application of these new tools to a 
set of two detectors, resulting in competitive limits obtained with a relatively modest exposure (section \ref{sec:results}).

\section{Detector response and energy thresholds}\label{sec:response}
\begin{figure}[htb]
\begin{center}
\includegraphics[width=1.0\textwidth]{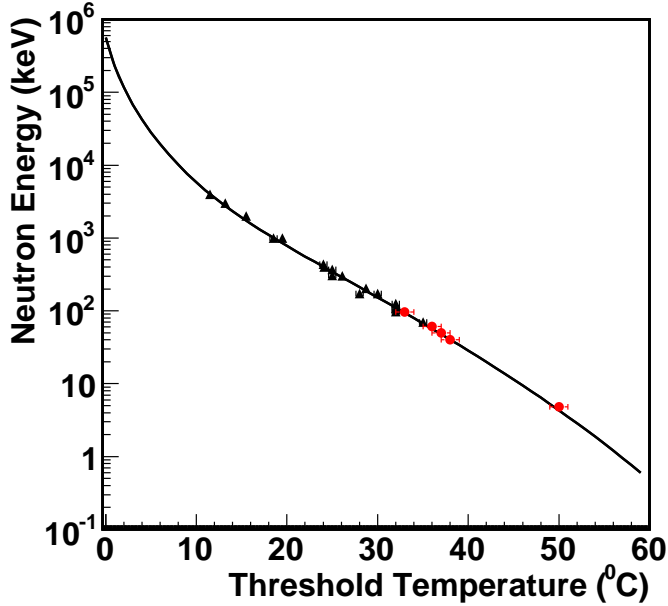}
\end{center}
\caption{Relation between neutron energy and threshold temperature for superheated C$_4$F$_{10}$ (1 bar). Mono-energetic neutrons were produced at the Montreal Tandem accelerator using the nuclear reactions $^{7}$Li(p,n)$^{7}$Be (black triangles) and $^{51}$V(p,n)$^{51}$Cr (red circles). In the case of $^{51}$V five resonances were selected yielding mono-energetic neutrons with sub-keV widths at E$_{n}$ = 4.8, 40, 50, 61 and 97 keV, respectively. Recoil energies of $^{19}$F at threshold are related to the neutron threshold energies by E$_{th}^F$(T) = 0.19E$_{th}^n$. Solid curve: prediction of the thermodynamical model.}\label{Fig:1}
\end{figure}

The active detector material in PICASSO is liquid C$_4$F$_{10}$, with a boiling temperature of T$_b$ = -1.7$^{\circ}$C 
at a pressure of 1.013 bar and a critical temperature of T$_c$ = 113.3$^{\circ}$C. The detector is formed by suspending droplets of C$_4$F$_{10}$ in an inactive polymerized gel matrix. The droplets are  
therefore in a metastable, superheated state at ambient temperature and pressure. In this condition a heat spike due to the energy deposited by an ionizing particle can cause the formation of a vapour bubble. This phase transition is explosive in nature and accompanied by an acoustic signal in the audible to ultrasonic frequency range. This can easily be registered by piezoelectric transducers.\\ 
In order for a phase transition to occur, a particle has to deposit a minimum amount of energy E$_{th}$ within 
a given length L$_{th}$.  Both, E$_{th}$ and L$_{th}$, decrease exponentially with temperature and are functions of the 
surface tension at the liquid-vapour interface and the superheat, which is defined as the difference between 
the vapour pressure of the liquid and the smaller external pressure. Details on the operation principle and the 
underlying thermodynamic model are described in \cite{boukhira00,PLB,NIM,seitz58}.\\               
If dE/dx is the mean stopping power of incident particles, then the energy deposited along L$_{th}$ is E$_{dep}$ = L$_{th}$ $\cdot$ dE/dx 
and the condition for triggering a phase transition becomes E$_{dep} \ge$  E$_{th}$.  The dependence of the threshold energy E$_{th}$ on temperature and pressure may be studied with neutron induced nuclear 
recoils. For this purpose extensive calibrations were performed at the Montr\'eal tandem accelerator facility. The mono-energetic neutrons used for calibration were produced via nuclear reactions with mono-energetic protons via the $^{7}$Li(p,n)$^{7}$Be and $^{51}$V(p,n)$^{51}$Cr reactions. The measurements with the Li target cover a range of neutron energies from 100 keV up to 5 MeV and the obtained results are discussed in detail in \cite{boukhira00,PLB}.  With improved proton beam stability at the Montr\'eal facility these calibrations could be recently extended with a $^{51}$V target down to a neutron energy of 4.8 keV.\\  
In order to acquire sufficient statistics close to threshold the proton beam energies were tuned to one of the numerous resonances in the $^{51}$V(p,n)$^{51}$Cr reaction cross section. In particular five resonances have been selected, each with a width smaller than one keV, yielding mono-energetic neutrons of E$_{n}$ = 4.8 keV, 40 keV, 50 keV, 61 keV and 97 keV \cite{gibbons}. The detectors used for these measurements were much smaller in size (30 mL) but fabricated in the same way, especially with similar droplet diameters, as the standard detectors used in dark matter runs. For each selected neutron energy data were taken by ramping the temperature up and down for a given pressure. Since close to threshold the cosmic ray induced n-background can amount to 50\% of the total count rate, each neutron run at a fixed temperature was followed by a background run at the same temperature. For a fixed neutron energy the data have been normalized by the integrated proton current and the count rate of a $^{3}$He counter mounted behind the target was used to compensate for short, off-resonance, beam energy fluctuations. For a given neutron energy the threshold temperature T$_{th}$ is then extracted by extrapolating the curves down to a few degrees below the lowest point measured.  The  relation between neutron energy and threshold temperature is shown in Fig. \ref{Fig:1}. 

\begin{figure}[htb]
\begin{center}
\includegraphics[trim=2cm 1cm 3cm 1cm,clip,width=.9\textwidth]{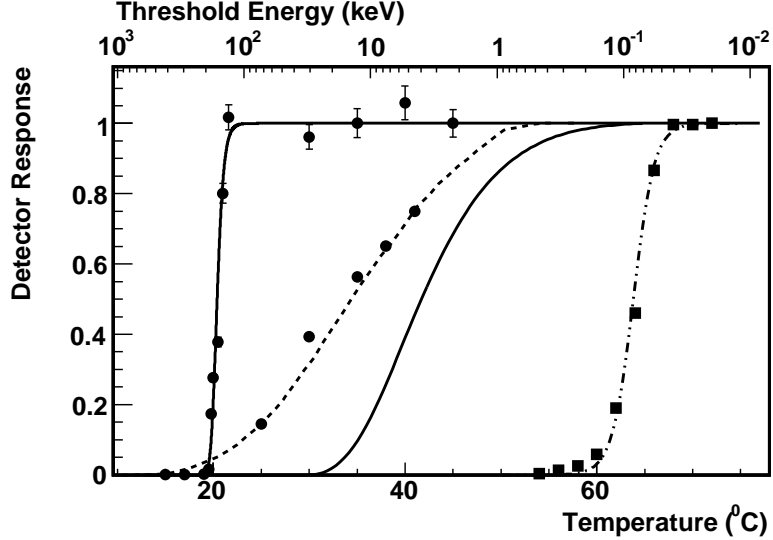}
\end{center}
\caption{Detector response to different types of particles as a function of temperature for 
detectors loaded with C$_4$F$_{10}$ droplets of $\sim$200$\mu$m in diameter. From left to right: 
alpha particles of 5.6 MeV in a detector spiked with $^{226}$Ra (fit to data points  represented 
by continuous line);  nuclear recoils from fast neutrons of an AmBe source compared to 
simulations (dotted line); expected response for nuclear recoils following scattering of 
a 50 GeVc$^{-2}$ WIMP (continuous line); response to 1.275 MeV gamma rays of a $^{22}$Na 
source (dashed line). All responses are normalized to one at full detection efficiency. 
Temperatures are converted into $^{19}$F threshold energies (upper x-axis) by using relation (\ref{eq:2}) 
in the text. If not visible, experimental errors are smaller than the symbols.}\label{Fig:2}
\end{figure}

For the energies interesting here, E$_{th}^n$(T)  follows an exponential temperature dependence. Outside this range thermodynamic considerations require that the threshold curve rejoins E$_{th}^n$(T$_{c}$) = 0 keV at the critical temperature T$_{c}$ and E$_{th}^n$(T$_{b}$) = $\infty$ keV at the boiling temperature T$_{b}$.  The energy of $^{19}$F recoils at threshold is then given by the relation: 
\begin{equation}\label{eq:2}
E_{th}^F(T) = 0.19E_{th}^n = (4.93\pm0.15)\times10^3 \exp\left[-0.173\times T(^{\circ}{\rm C})\right] {\rm (keV)}
\end{equation}
where the kinematic factor 0.19 relates the measured neutron energy to the corresponding maximum fluorine recoil energy. 
The error of 3\% is largely due to the systematic errors in the temperature measurements during the test beam runs.\\ 
The measurements at the lowest neutron energy (E$_{n}$ = 4.8 keV or E$_{th}^{F}$ = 0.9 keV) are particularly challenging since at threshold and above, the detector had to be operated between 48$^{\circ}$C and 60 $^{\circ}$C where C$_4$F$_{10}$ becomes 
sensitive to the 320 keV gamma rays (T$_{1/2}$ = 28 d) following de-excitation of $^{51}$Cr. Therefore this background had to be measured independently during a beam off period after each neutron run and subtracted from the data. Since the detector is not sensitive to gamma-rays themselves, but rather to $\delta$ -electrons with energies smaller than a few keV produced on the tracks of Compton scattered electrons \cite{NIM}, the observed gamma sensitivity affirms the very low energy thresholds attained at 50$^{\circ}$C (see also Fig. 2).
The temperature range of operation in PICASSO with 18.5$^{\circ}$C $<$ T $<$ 45$^{\circ}$C then translates (in reversed order) into sensitivities to $^{19}$F recoils with energies in the range of 2.0 keV $<$ E$_F$ $<$ 200 keV.
Since each temperature corresponds to a well defined recoil energy threshold, the spectrum of the particle induced energy depositions can then be reconstructed by ramping the temperature.\\ 
It was found moreover that the detection threshold is not a sharp step function, but rises gradually from threshold to full efficiency \cite{derrico01}. The probability P(E$_{dep}$, E$_{th}$), that an 
energy deposition E$_{dep}$ larger than the energy threshold E$_{th}$ will generate a nucleation can be approximated by:
\begin{equation}\label{eq:3}
P(E_{dep},E_{th}(T))=1-\exp\left[\alpha\left(1-\frac{E_{dep}}{E_{th}(T)}\right)\right]
\end{equation}
where the parameter $\alpha$ describes the observed steepness of the threshold. This parameter is related to the intrinsic energy resolution and reflects the statistical nature of the energy deposition and its conversion into heat; it has to be determined experimentally for each superheated liquid and for different particle species, respectively.  The adopted value of $\alpha = 2.5 \pm 0.5$ corresponds to the best fit to the temperature response functions where the resonances were best defined. This new value is larger than our earlier quoted result of $\alpha = 1.0 \pm 0.1$ \cite{PLB, NIM}; this is explained  by the fact that now the $\alpha$-parameter has been re-evaluated using simulations which take into account the low energy tail in the neutron beam profile due to scattering of neutrons off the environment of the detector area. It is thus normal that $\alpha$ should have a larger value than when it was assumed that the neutron beam was a delta function. The new adopted value also reproduces best poly-energetic neutron source spectra and (n$_{th}$, p) - response functions for chlorine containing freons \cite {derrico01,faust08,auger08}. Work is in progress to still better define this parameter and in particular to investigate a suspected energy dependence.\\ 
	Complete simulations of the detector response to mono-energetic test beam neutrons  and poly-energetic neutrons from an AcBe source show that all data are in good  agreement with the simulations and can be well described with a consistent set of the variables which parametrize the underlying model assumptions \cite{NIM}. The 
response to different kinds of particle interactions depends on the respective specific energy losses. Examples are shown in Fig.\ref{Fig:2}: WIMP induced recoil energies of $^{19}$F 
nuclei  are expected to be smaller than 100 keV and therefore become detectable above 30$^{\circ}$C. 
Particles which produce only low ionization densities, such as cosmic ray muons, $\gamma$ - and $\beta$ - rays are observed at temperatures above 50$^{\circ}$C. They are well separated from the strongly ionizing neutron or WIMP induced recoils, which allows an efficient suppression of such backgrounds at the level of 10$^{-8}$ to 10$^{-10}$ \cite{PLB}.\\  
	Alpha particles show a different behaviour. Data obtained with $^{226}$Ra spiked detectors and detector modules with a significant contamination of U, Th, or their daughters show that the sensitivity to alpha particles, which are contained entirely in the droplets, starts with a sharp step at 21$^{\circ}$C (Fig. \ref{Fig:2}).
At higher temperatures the liquid becomes sensitive to smaller dE/dx on the tracks and the recoiling daughter nucleus itself. However, since the detector is already fully sensitive to alpha particles immediately above threshold, the temperature response levels off in a plateau and the detector remains fully sensitive to alpha particles 
over the entire range of the WIMP sensitivity. Therefore, since the shapes of the WIMP and alpha curves differ substantially, the alpha background can be efficiently discriminated by measuring the temperature profile of the detector response. This 
property is exploited in this analysis (section \ref{sec:disc}).  
\section{Detectors and experimental set up}\label{sec:det}
The current generation of PICASSO detectors consists of cylindrical modules of 14
cm inner diameter, 40 cm height and a wall thickness of 1.4 cm. They are fabricated 
from acrylic and are closed on top by stainless steel lids sealed with polyurethane 
O-rings. Each detector is filled with 4.5 litres of polymerised emulsion loaded with 
droplets of C$_4$F$_{10}$. The active part of each detector is topped by a 4 cm thick 
layer of mineral oil, which is connected to a hydraulic manifold.
	The typical distribution of the droplet sizes seen by neutrons or WIMPS (i.e. weighted by droplet volume) peaks at a diameter of  200 $\mu$m with a variance of around 100 $\mu$m. The mass of C$_4$F$_{10}$ in each detector is about 85 g, corresponding to 68 g of $^{19}$F. It can be determined with a precision of 1\% by weighing the detector components during the fabrication. Additional uncertainties in the mass determination are attributed to inactive C$_4$F$_{10}$, either due to diffusion of C$_4$F$_{10}$ into the gel ($<$ 1.7\%) or contained in microscopic droplets which might not be able to trigger ($<$ 1\%). Therefore after fabrication the detector loading is verified by a  separate measurement in the known flux of an AcBe neutron source with a combined statistical and systematic error of 5\%. Both methods were found to yield consistent results within the given uncertainties. The active mass quoted for each detector is then the one determined 
by weighing to which an uncertainty of 5\% is assigned.\\  
	The present installation at SNOLAB accommodates 32 detector modules. A group of four detectors is installed on rubber shock absorbers in a thermally and acoustically insulated box, serving as a temperature control unit. The external dimensions are 56 x 65 x 65 cm$^3$. These sub-units are arranged into a larger, eight unit super-cube. The operating temperature in 
each sub-unit can be varied independently from 20$^{\circ}$C to 55$^{\circ}$C and the temperature is 
regulated in each unit individually with a precision of  $\pm$ 0.1$^{\circ}$C.  Finally the entire 
installation is surrounded by 238 water cubes of 30.5 cm edge-length, which serve as neutron 
moderator and shielding.\\
	At the location of the experiment at SNOLAB, with a depth of 2070 m, the cosmic muon flux is reduced to a level of 0.29 muons m$^{-2}$ d$^{-1}$ \cite{SNO09}.  The ambient thermal neutron flux from the rock was measured to be 4144.9 $\pm$ 49.8 $\pm$ 105.3 neutrons m$^{_-2}$ 
d$^{-1}$ \cite{Browne}. The fast neutron flux is less well known, but has been estimated to be about 4000 neutrons m$^{-2}$ d$^{-1}$. Since the entire set up is tightly surrounded by water cubes, 
which provide a 30.5 cm thick shielding against fast neutrons coming from the rock, we expect from Monte Carlo simulations another factor 100 in flux reduction. This would result in an 
expected count rate at the level of 10$^{-5}$ counts h$^{-1}$ g$^{-1}$, which is two orders of magnitude below the count rate of the detectors which are the subject of this analysis.\\
	After a measuring cycle of 40 h at ambient mine pressure (1.2 bar, unregulated) the detectors are recompressed for 15 h at a pressure of 6 bar in order to reduce any C$_4$F$_{10}$ gas bubble back to a droplet in order to prevent further bubble growth or coalescence which could damage the polymer. For detector recompression, each unit is equipped with a hydraulic pressure manifold serving four detectors. The pressure is continuously monitored and controlled.\\ 
	Each detector is read out by nine piezoelectric transducers on  the outside wall of the containers at three different heights. The middle level is rotated  by 60$^{\circ}$  with respect to the other two levels to provide a uniform distribution of sensors. The transducers are constructed using ceramic disks (PZ27 Ferroperm) 
with a diameter of 16 mm and 8.7 mm in thickness. A typical transducer signal starts with a fast rise, a maximum typically within the first 100 $\mu$s, followed by a series of slower oscillations, which settle 
down after several milliseconds. The signals are amplified by high impedance pre-amplifiers with a gain of 3500, resulting in a sensitivity for pressure amplitudes of 1000 mV Pa$^{-1}$. High and low 
pass RC filters in the front end electronics define  
a band pass concentrated between 10 kHz and 50 kHz. 
The lower limit was chosen in order to reduce low frequency acoustic noise and the upper limit was imposed by the timing requirements of the ADC system \cite{gornea07}.\\
	An event triggers if at least one transducer signal is larger than a threshold of 300 mV and all nine transducer channels of a detector are read out. The trigger is  fully efficient for 
signals above 24$^{\circ}$C, which is the temperature region of concern in this analysis. At temperatures below those used in this analysis the decreasing signal amplitudes start to fall below threshold. In this case the trigger efficiency is inferred from the truncated signal amplitude distributions. The dead time of the DAQ system is negligible during normal data runs. By comparing the signal arrival times of the different sensors, the position of each event can be reconstructed in 3D with a precision of  $\pm$ 5 mm. This will allow the fiducialization of events in future analyses.              		

\section{Event by event discrimination and analysis}\label{sec:disc}
\begin{figure}[htb]
\begin{center}
\includegraphics[trim=2cm 2cm 3cm 3cm, clip, width=.45\textwidth]{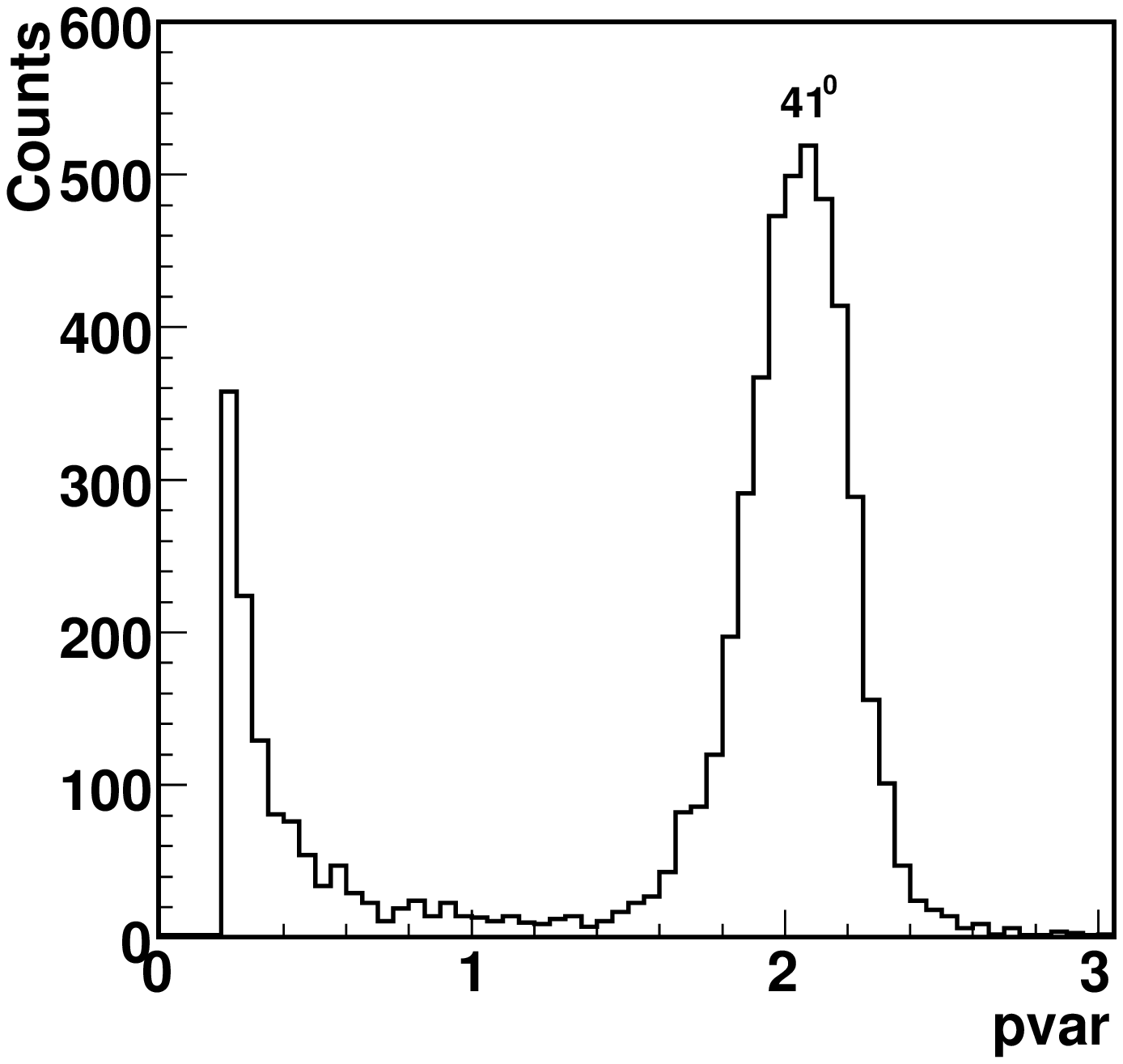}
\includegraphics[trim=2cm 2cm 3cm 3cm, clip, width=.45\textwidth]{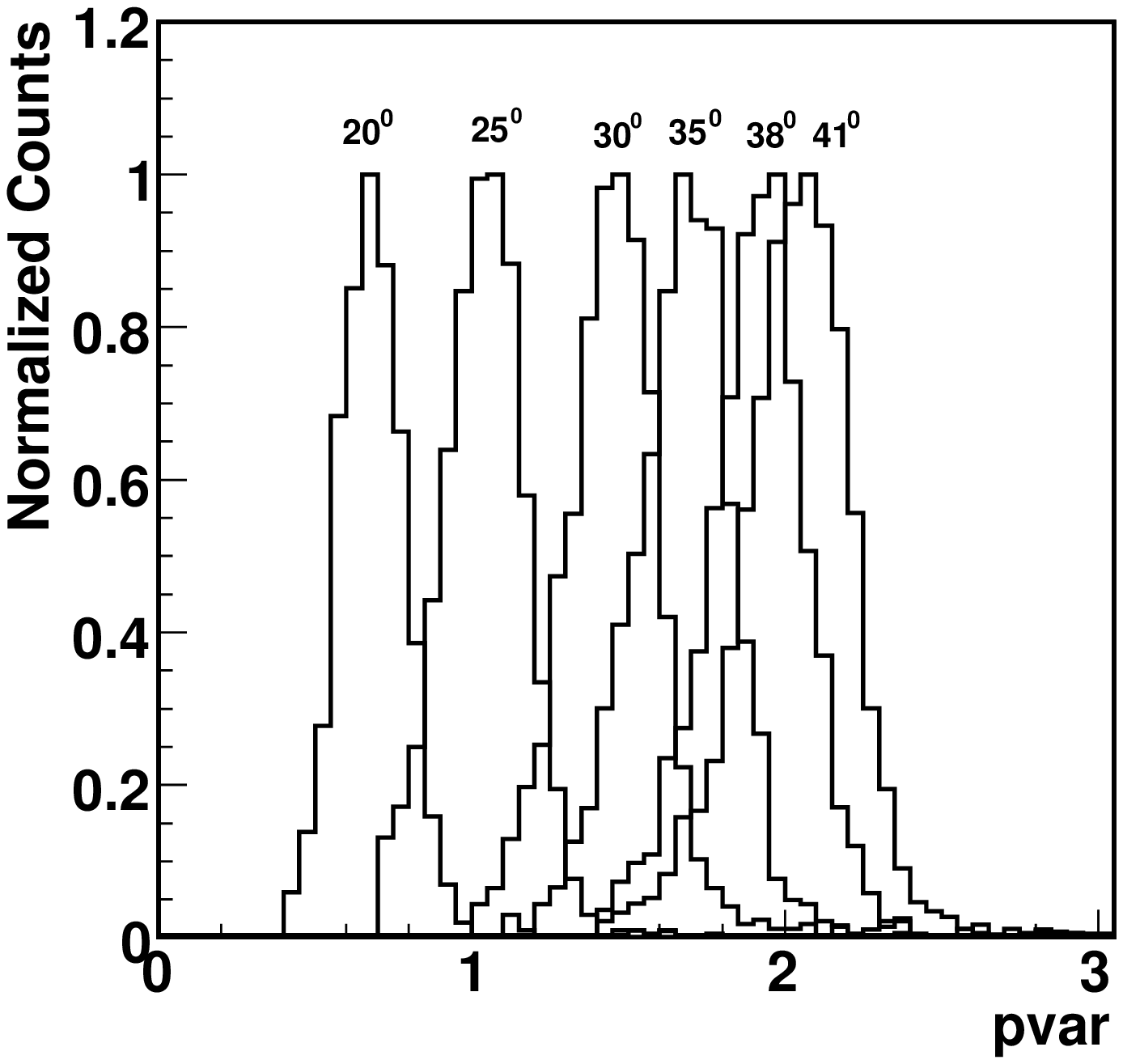}
\end{center}
\caption{Distribution of the integrated signal power pvar recorded in calibrations with poly-energetic neutrons from an AmBe source. For a given event {\it pvar} is constructed by averaging over the waveforms of at least six transducers per detector. Neutron induced recoils
accumulate in a peak (the same peak where WIMP induced recoils are expected); this peak is well separated from acoustic and electronic noise (left). The measured temperature dependence of the recoil peaks (right) serves to define a temperature dependent cut (95\% acceptance) to the left of the centroid in order to reject non-particle induced events.}\label{Fig:3}
\end{figure}

\begin{figure}[htb]
\begin{center}
\includegraphics[trim=1.5cm 1.7cm 3cm 2.5cm, clip, width=.85\textwidth]{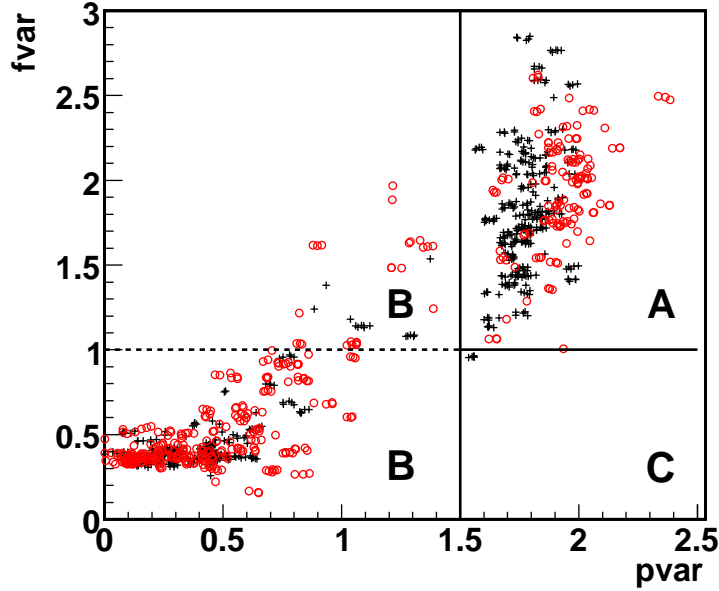}
\end{center}
\caption{The signal energy and frequency related variables pvar and fvar allow the classification of events into distinct categories. Shown are data from n- calibration (black crosses) and WIMP runs (red circles) for det. 71 used in this analysis. Neutron induced nuclear 
recoils from AmBe calibrations are located in region A; alpha particle events from WIMP runs appear displaced to the right in region A; electronic and acoustic noise, such as mine blasts, populate the lower region B; fracture events produced by primary events in the polymer would be located in region C, but do not occur in this detector. Data were recorded at 35$^{\circ}$C.}\label{Fig:4}
\end{figure}

In addition to passive shielding against neutrons and gamma rays, dark matter experiments usually exploit specific features in the transients of the primary signal,  such as pulse shape, or rise or 
decay times, in order to discriminate between nuclear recoils and other events. A search in PICASSO for signatures in the recorded transducer signals showed that the intensity and frequency content of the acoustic signals indeed contains information about the nature of 
the primary event \cite{aubin08}. From this information, variables can be constructed, which allow an event by event discrimination between particle and non-particle induced backgrounds.\\ 
	 Calibration data with fast neutrons of AcBe, AmBe and Cf sources showed that the amplitude distributions of the high frequency content of the signals ($>$18 KHz) have a well defined peak.  In addition it was found that alpha particle induced signals are significantly more intense than those of neutrons (and therefore of WIMP events).  In 
both cases the energy released during vaporisation increases with temperature, but stays well defined for a given temperature. A detailed discussion of this phenomenon can be found in \cite{aubin08}.\\
	The variable {\it pvar} is a measure of the integrated sound intensity of the fast component of an event signal. In order to construct {\it pvar}, first a Bessel band pass filter is used to cut off frequencies below 18 kHz. For each event, the recorded 
waveform from each transducer is squared and integrated over the signal duration, starting from a fixed pre-trigger time.  The logarithm of this integral is taken, and the resulting values are averaged over all active transducers for each event. Finally the averaged value 
is rescaled such that {\it pvar} = 0 corresponds to the lowest observed value in each detector \cite{auger08,gornea08,giroux08}. Particle induced events, such as neutrons and alpha particles, show up as well defined distributions, clearly separated from noise (Fig. \ref{Fig:3}) and the centroid varies smoothly as a function of temperature.\\
	Another acoustic signature for particle induced nucleations has been constructed from the power spectrum of the fast Fourier transforms (FFT) \cite{clark08}. Studies of the FFT of neutron induced signals have shown that the majority of the signal power resides in a 
frequency range from 20 kHz up to 70 kHz. By taking the logarithm of the ratio of the contents within the intervals of 20 kHz up to 30 kHz and 45 kHz up to 55 kHz, another variable, {\it fvar}, is constructed which also allows the discrimination of particle induced and non-particle induced events.\\ 
	A scatter plot of {\it fvar} vs. {\it pvar} at 35$^{\circ}$C is shown in Fig. \ref{Fig:4} and various regions of different signals and backgrounds are identified. Particle induced events such as 
neutron induced recoil nuclei from calibration runs with poly-energetic neutrons and background events obtained during WIMP runs are concentrated in region A. The neutron induced recoils are 
well defined in a band between  1.6 $<$ {\it pvar} $<$ 1.8, whereas events caused by alpha particles extend to larger values (1.6 $<$ {\it pvar}  $<$ 2.3).  Acoustic signals which fall into region B 
are for example blast events created by mining activity. 
Also in region B are events due to sporadic elctronic pulses with amplitudes larger than 300 mV trigger the system. The lack of these events in the preceding DAQ system indicates that these events are electronics related. The corresponding transients in other channels
have much smaller amplitudes probably due to electronic cross talk. These events show up at very small values of {\it pvar} and are completely eliminated by the applied cuts. Another class of rare events was observed wherein two events occur within a time separation of 0.01 s. This corresponds to the minimum time difference allowed by the data 
acquisition system for a subsequent event to be recorded. In this case the data acquisition is re-triggering on the long lasting ringing of a primary signal and produces a false event.  
These events are removed by the {\it pvar}, {\it fvar} cuts and an additional time cut discussed below.\\
	The so-called fracture events, occur when a primary event causes  several secondary events in the same area. These would appear in region C. Since the polymer matrix can be weakened or fractured, some secondary events might appear several seconds after the primary event. This effect has been studied by exposing a test detector to a high neutron flux in order to enhance the occurrence of this kind of events. It was found that the follow-up events which are not due to particle induced nucleations have a different response in the frequency variable {\it fvar}; they occur in region C and are efficiently removed.\\ 	
	These discriminating variables have been applied first to a set of two detectors (71 and 72) with $^{19}$F target mass of 65.06 $\pm$ 3.2 g and  69.0 $\pm$ 3.5 g, respectively. The analysis is presently being extended to the remaining 30 detectors, which were installed between late 2008 and 2009 and we expect a very competitive limit will be obtained when the detectors have been fully characterized and the analysis can be completed. The runs used for this analysis covered the period from June 2007 until July 2008, with three neutron source calibration periods interspersed, which allowed monitoring the stability over the period of data taking. The temperatures were varied in the range from 45$^{\circ}$C  to 18.5$^{\circ}$C, which corresponds to a range of threshold energies from 2 keV to 200 keV. A total of  101.5 d  (103.5 d)  of data taking time has been used in this analysis for det. 71 (72), respectively, which results in a total exposure of 13.75 $\pm$ 0.48 kgd.  Two independent analyses have been performed and yielded consistent results.\\

The analysis proceeds in 4 steps:
\begin{enumerate}[a)] 
\item In analysing a WIMP run, at least six working acoustic channels per detector are required and the first hour of each run is removed in order to allow for detector stabilization after re-pressurizing and de-pressurizing (1hr-off cut). 
\item Next, a cut is applied on the power variable {\it pvar}. In order to determine the best value of the {\it pvar} cut to use, neutron calibration runs were carried out at defined temperatures with a weak AmBe neutron source (69 s$^{-1}$), placed 10 cm, equidistant from the detectors. The results are shown in Fig. \ref{Fig:3} for one of the detectors (det. 72), used in this analysis. Once the {\it pvar} distributions are obtained from calibration runs, a Gaussian curve is fit to the neutron response curve and a cut value, pcut (T) 
is determined such that 95\% of neutron (or WIMP) induced recoils with
{\it pvar} $>$ pcut (T) are accepted.  For temperature points where no calibrations have been done, cut values are inferred by interpolating the curve fit to the measured pcut (T) values. In a given run, all events with {\it pvar} $>$ pcut (T) are selected as particle induced bubble events.\\
	Gain instabilities, variations in acoustic signals and variations in material properties that might give rise to a shift of the signal amplitudes and which could affect the stability of {\it pvar} over time were studied. In particular a gain map was established for all transducer channels from the neutron calibration data and the 
average signal amplitudes of each channel was adjusted to a common reference value. These adjustments were found to be smaller than $\pm$ 7\% per channel and had a negligible influence on the analysis. Repeated neutron calibrations showed that changes of the {\it pvar}-distributions were smaller than 2.5\% over the period under consideration.  
\item A wide cut on the frequency variable {\it fvar} is applied with an acceptance of $>$ 98\% for particle induced events, mainly in order to remove spurious blast or fracture events which could 
contaminate the signal region. This cut does not vary with temperature.  
\item A time veto is applied after each event: in normal data runs signal rates should ideally be due to WIMPs, or background events due to ambient neutrons, gammas or alpha particles from U/Th contaminations and their daughters. All these are randomly produced and as such, a distribution of the time intervals between successive events, $\Delta$t, should follow Poisson statistics with the mean being equal to 
the inverse of the count rate  of the associated background source. Therefore, in order to check the quality of our data, after application of pcut (T)  the Poisson expectation is fit to the $\Delta$t distributions in all of our WIMP runs and the reduced chi square for 
different $\Delta$t cut values is checked. We noticed, that before applying the {\it pvar} cut, an excess of events with time differences smaller than 200 s appeared (fracture events; see section \ref{sec:det}). Although after application of pcut no statistically relevant degradation of the chi-squared was observed, a precautionary veto of $\Delta$t = 200 s was applied for both detectors at the cost of an increase in dead time by 3.7\% and 1.1\% for det. 71 and det. 72 respectively.\\
	In summary, four cuts were used: the temperature dependent pcut (T), with an acceptance of 95\% for particle induced events, a wide cut in the frequency variable {\it fvar}, a 1hr-off cut after decompression and a $\Delta$t = 200 s veto between successive events. The effects of  these cuts are shown in \ref{tab:1}. The most selective cut is pcut(T) which rejects 75\% and 90\% of all triggers for det. 71. and 72, respectively. 
\end{enumerate}

\begin{table}
\centering
\begin{tabular}{ccc}
\hline
&\multicolumn{2}{c}{Number of events remaining after each cut in sequence.}\\
&	Detector 71	&Detector 72\\
\hline
Total triggers	&7322	&5784\\
After pcut / fcut & 	1836&	566\\
After 1hr-off cut &	1768&	543\\
After  $\Delta$t cut&	1555&	496\\
\hline
\end{tabular}
\caption{Effect of  applied cuts. Since  the observed count rates are 
constant over the observed temperature range, all WIMP runs have been combined together.}\label{tab:1}
\end{table}

\section{Results}\label{sec:results}
\begin{figure}[tb]
\begin{center}
\includegraphics[trim=2cm 1cm 3cm 2cm, clip, width=.8\textwidth]{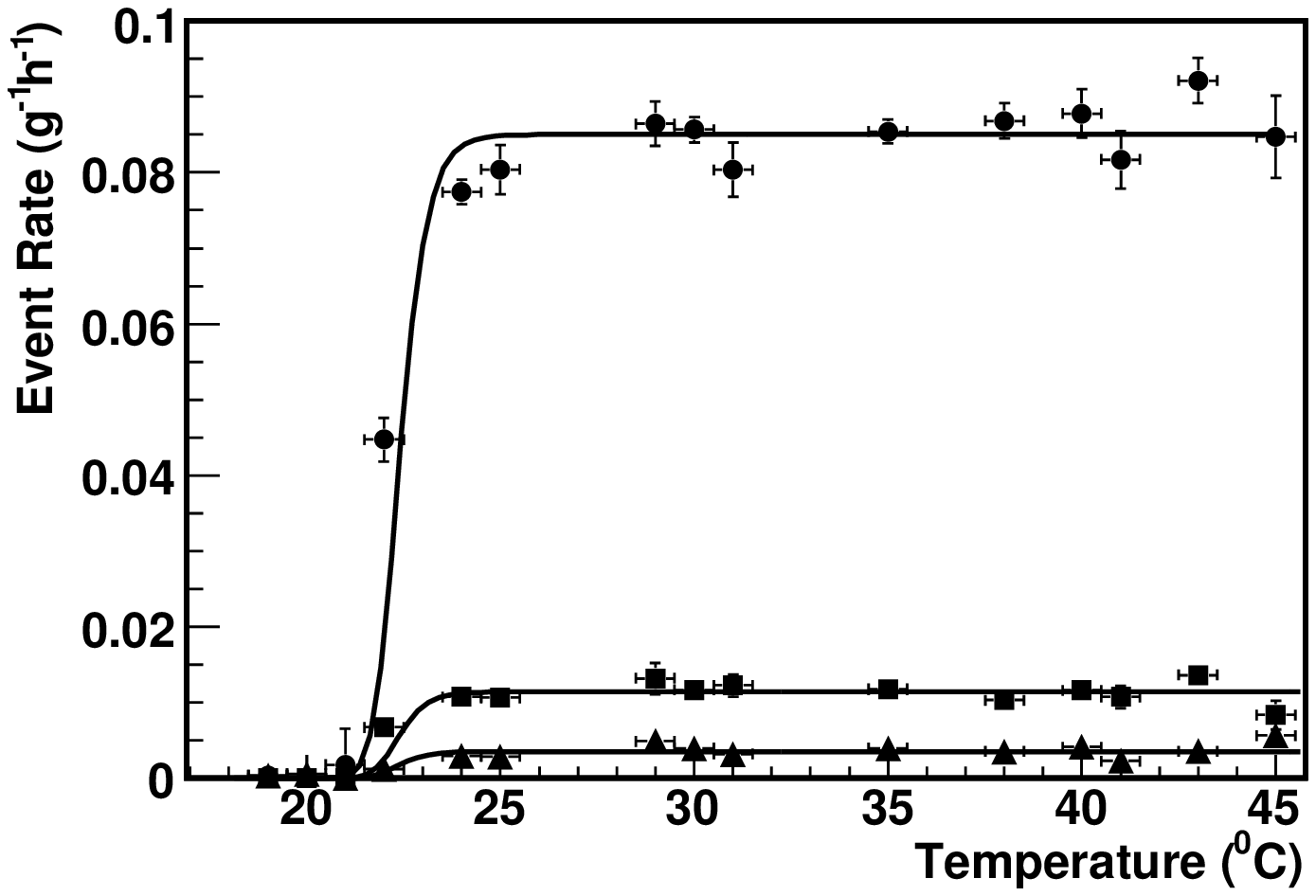}
\includegraphics[trim=2cm 1cm 3cm 2cm, clip, width=.8\textwidth]{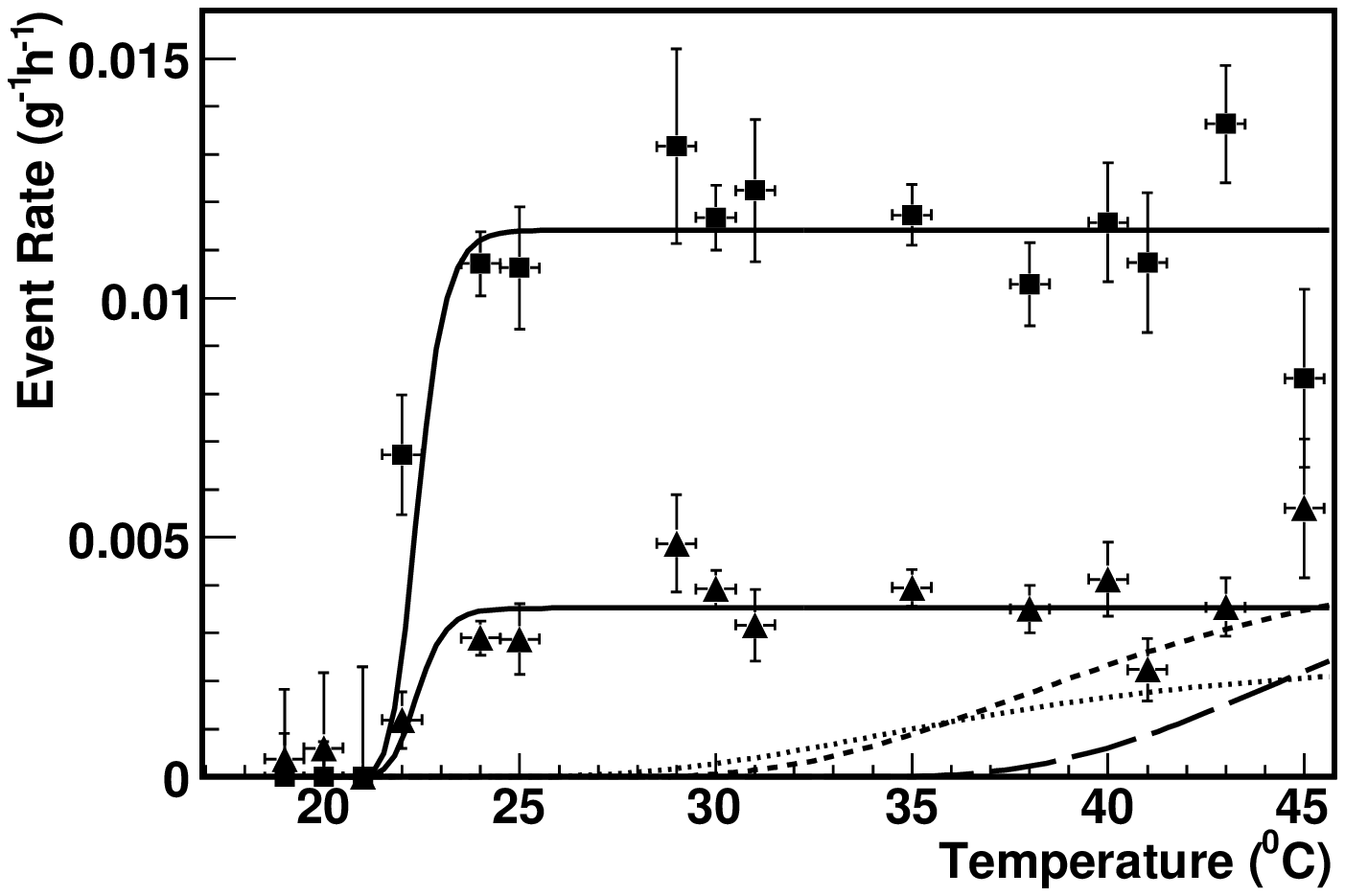}
\end{center}
\caption{Count rates as a function of temperature. The rates are normalized by the active mass of $^{19}$F and indicate the different levels of $\alpha$-background. Top: detector 93 (dots) with its higher $\alpha$-background serves as a reference to define  
the $\alpha$-threshold; detectors 71 (squares) and 72 (triangles) are used in this analysis. Bottom: zoom of the rates of det. 71 and 72. Also shown are the expected response curves for WIMP induced nuclear recoils for M$_W$ = 10 GeV c$^{-2}$ (broken), 30 GeV c$^{-2}$  (dashed)  and 100 GeV c$^{-2}$  (dotted); a cross section of  $\sigma_p$ = 1 pb  was assumed for clarity. }\label{Fig:5}
\end{figure}

After correction for dead times and the {\it pvar} acceptance the
count rates are  normalised by the amount of target mass in each detector (i.e. grams of  $^{19}$F) and the  respective live times in order to obtain the count rate in counts h$^{-1}$ g$^{-1}$ as a function of temperature, as shown in Fig. \ref{Fig:5}.  The count rates of detectors 71 and 72 show a steep threshold at 21.5$^{\circ}$C and a flat plateau between 24$^{\circ}$C and 45$^{\circ}$C. 
At higher temperatures the count rates increase again sharply, when
the detectors start to become sensitive to delta electrons produced on
the tracks of Compton electrons. The scattering gamma rays are produced in the detectors themselves and are mainly due to  $^{40}$K which is a contaminant of the CsCl salt used in the fabrication.

	In the plateau region the averaged count rates are 0.0114 $\pm$ 0.003 counts h$^{-1}$ g$^{-1}$ for det. 71 and 0.0037 $\pm$ 0.0002 counts h$^{-1}$ g$^{-1}$ for  det. 72. In Fig. \ref{Fig:5} (top) the two data sets are compared with those of detector 93, with a similar loading and droplet size distribution, but with higher background count rate due to U/Th contaminations. All three detectors exhibit temperature profiles identical to the response of 
the $^{226}$Ra spiked detector shown in Fig. \ref{Fig:2} and the observed count rates are attributed to alpha particles. The thresholds are shifted by 1.5$^{\circ}$C to higher temperatures with respect to the  $^{226}$Ra-data, which is expected because of  the increased ambient pressure of 1.2 bar at SNOLAB. In order to quantify this shift in threshold we took advantage of the improved statistics of detector 93. A fit of the alpha response curve, determined from the $^{226}$Ra spiked detector to these data results in a reduced chi-squared of 1.15, 1.25 and 1.5 for the detectors 71, 72 and 93, respectively. The alpha nature of the recorded  events has been confirmed independently by applying the alpha-neutron discrimination feature discussed in section \ref{sec:det} and in more detail in \cite{aubin08}. In the very probable case that the alpha emitters are located in the polymer matrix we expect a geometric detection efficiency for alpha particles of 0.3\%. This would result in an alpha activity in det. 72 of 3.3 mBq 
per kilogram of total detector material, which translates into a contamination 
at the level of 2.7 x 10$^{-10}$ gU g$^{-1}$ or 8.1x10$^{-11}$ gTh g$^{-1}$, if 
secular equilibrium is assumed and if only one isotope contributes.\\ 
	In order to compare the data of detectors 71 and 72 to the 
expected $^{19}$F recoil spectrum for interactions with WIMPs in our galactic halo, we follow the recommendations in \cite{levin96} and use as halo parameters a local WIMP matter density of 0.3  GeVc$^{-3}$, a neutralino velocity dispersion in the halo of 230 km s$^{-1}$, an earth-halo relative velocity of 244 km s$^{-1}$ and a 
galactic escape velocity of 600 km s$^{-1}$. Keeping the WIMP mass and the cross section as parameters, the expected WIMP induced count rate can be  calculated as a function of temperature by applying the threshold function given in (\ref{eq:3}) 
to the expected WIMP induced recoil spectrum. Details of these calculations can be found in \cite{PLB,NIM}. The resulting response curves are shown for several WIMP masses in Fig. \ref{Fig:5} (bottom). They differ significantly in shape from the flat alpha background and therefore by fitting the two distributions to the data an upper 
bound can be set on the interaction cross section $\sigma_F$  on $^{19}$F for a given value of  WIMP mass M$_W$.
\clearpage
The two free parameters of the fit are the  scale factor of the alpha plateau and the cross section $\sigma_F$ (M$_W$)  for a given WIMP mass. The minima of the reduced chi-squares were found to lie around $\chi^2$ = 1.2 and 1.1 for detectors 71 and 72, respectively over a mass range from M$_W$ = 7 to 1000 GeVc$^{-2}$. By combining the fit 
results of both detectors in a weighted average we obtain the maximum sensitivity for a WIMP mass of M$_W$ = 24 GeVc$^{-2}$ and a cross section of  $\sigma_F$ = - 0.446 pb $\pm$ 10.83 pb $\pm$ 0.61 pb (1$\sigma$), compatible with no effect. This result can be converted into a limit on the cross section for WIMP induced reactions on $^{19}$F of $\sigma_F$  = 13.9 pb  (90\% C.L.).\\	
	The following systematic uncertainties (1$\sigma$) have been included in the evaluation of these limits: (a) a 20 \% uncertainty in the energy resolution parameter $\alpha$ resulting in a  2 \% error on the cross section limits; (b) a 5 \% error in the determination 
of the active mass of the detectors (see section \ref{sec:response}); (b) a 3 \% error in the energy scale due to systematic errors in the temperature measurements during test beam calibration and which lead to an uncertainty of 2 \% on the cross section limits; (c) a 2.5 \% uncertainty in the pvar cut acceptance and a 1.5 \% uncertainty due to curve fitting of pcut(T) result in a 3 \% uncertainty in the limits; (d) atmospheric pressure changes at the level of 3 \% introduce uncertainties in the superheat and are equivalent to a corresponding uncertainty in the temperature scale of  $\pm$ 0.4$^{\circ}$C at 20$^{\circ}$C and  $\pm$ 0.1$^{\circ}$C at 40$^{\circ}$C; the 
latter result in an  uncertainty of about 1\% on the limits; (e) similarly the hydrostatic pressure gradient of $\pm$ 2\% with respect to the centre of each detector module is translated  into a temperature uncertainty over the detector and introduces another 
uncertainty of $\pm$ 1\% in the cross section.\\                                                                          
	In order to compare our results to other experiments we follow the model independent procedure outlined in \cite{tovey,giulianiPRL04,giulianiPLB04} and express our results in terms of WIMP scattering on nucleons, by assuming that all events 
are either due to WIMP-proton scatterings (a$_n$ = 0) or WIMP-neutron scatterings (a$_p$ = 0). Then the cross section on the nucleons can be written as: \begin{equation}\label{eq:4}
\sigma_{n,p} = \sigma_F \left(\frac{\mu_{p,n}}{\mu_F}\right)^2 \frac{C_{p,n}}{C_{p,n(F)}}
\end{equation}
where $\mu_{p,n}$ are the WIMP-nucleon reduced masses and C$_{p,n}$  are the enhancement factors for scattering on free nucleons. The enhancement factor for nucleon scattering on $^{19}$F is given by C$_{n,p(F)}$  = (8/$\pi$)a$_{p,n}^2<$S$_{p,n}>^2$ (J+1)/J  
and the values of the ratios C$_{p(F)}$ / C$_p$ = 0.778 and C$_{n(F)}$ / C$_n$  =0.0475 are taken from \cite{pachenko89}. 
\begin{figure}[htb]
\begin{center}
\includegraphics[width=.9\textwidth]{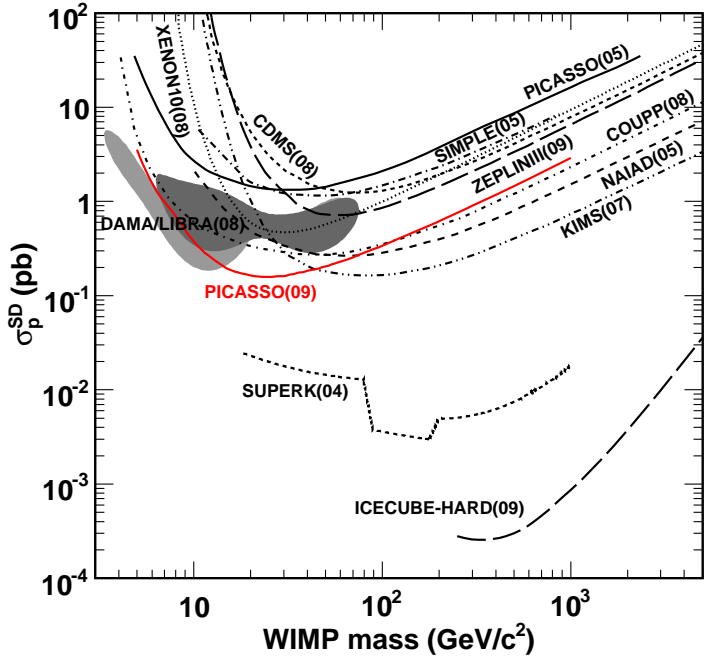}
\end{center}
\caption{Upper limits on direct spin-dependent WIMP-proton cross sections. PICASSO limits are shown as full lines. Additional curves are COUPP \cite{behnke08}, 
ZEPLIN \cite{lebedenko09}, SIMPLE \cite{girard05}, CDMS \cite{ahmed09}, KIMS \cite{lee07}, NAIAD \cite{alner07}, XENON10 \cite{angle08}. The observed annual modulation by the DAMA/LIBRA \cite{bernabei} experiment interpreted by \cite{savage} in terms of spin-dependent cross sections is shown as the filled regions (dark grey: without ion 
channelling, light grey with ion channelling). Also included are the limits from the indirect searches of Super-Kamiokande \cite{desai04} and IceCube \cite{abasi09}. Data courtesy of \cite{gaitskell}.}\label{Fig:6}
\end{figure}

\begin{figure}[htb]
\begin{center}
\includegraphics[width=.9\textwidth]{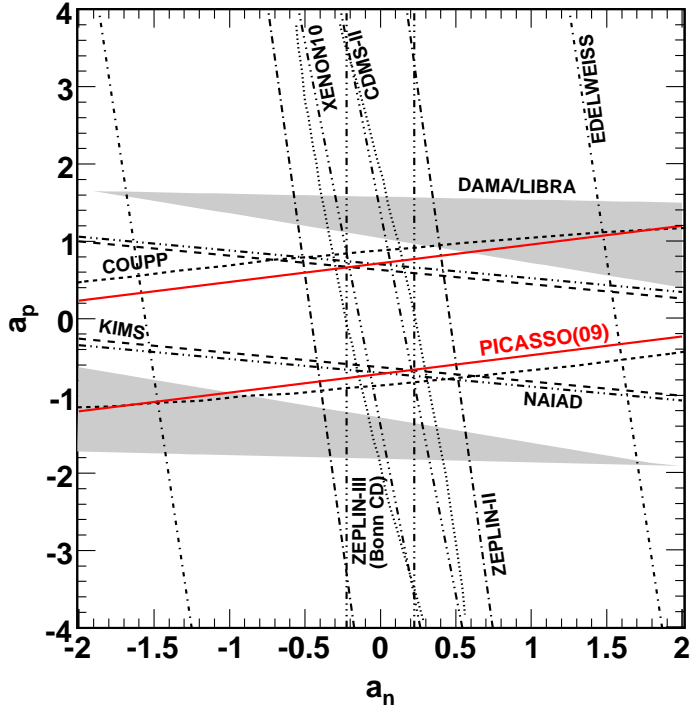}
\end{center}
\caption{Plane of the effective coupling strengths a$_p$, a$_n$ for 
protons and neutrons for a WIMP mass of 50 GeVc$^{-2}$. The allowed 
DAMA/LIBRA region discussed in \cite{savage} is indicated as the  
grey shaded region. Each other experiment excludes the exterior of 
the corresponding bands at 90\% C.L. This plot has been adopted 
from \cite{lebedenko09} and refers to the references given in Fig. \ref{Fig:6}.}\label{Fig:7}
\end{figure}

	Using (\ref{eq:4}) we can translate the obtained fit results for $\sigma_F$ into a cross section measurement on free protons of  $\sigma_p$ = - 0.0051 pb $\pm$ 0.124 pb $\pm$ 0.007 pb (1$\sigma$) 
which results in a limit of $\sigma_p$ = 0.16 pb (90\%C.L.) for a WIMP mass of 24 GeVc$^{-2}$. The resulting exclusion plot for the cross section on protons as a function of the WIMP mass is shown in Fig. \ref{Fig:6}, where we also compare the obtained  limits with the most recent results of other experiments in the spin-dependent sector.\\ 	 
Since the nuclear spin in $^{19}$F is carried essentially by the 2s$_{1/2}$ proton, the sensitivity to neutrons is small, but not negligible. Adopting the nuclear 
model for $^{19}$F discussed in \cite{pachenko89}, the sensitivity to WIMP scattering on neutrons should be reduced by a factor 0.061 with respect to protons. This leads to a cross section limit on free neutrons of $\sigma_n$ = 2.60 pb at 90\% C.L. (M$_W$ = 24 GeVc$^{-2}$ ), which is about a factor of 300 less stringent than the leading constraints for WIMP neutron scattering obtained by CDMS \cite{ahmed09}, XENON10 \cite{angle08} and ZEPLIN \cite{lebedenko09}.\\   
	Once the limits on the free nucleon cross sections are obtained, the allowed regions in the parameter plane of the nucleon coupling strengths a$_p$ - a$_n$ can be found following the prescriptions in \cite{tovey,PLB}. In PICASSO  the allowed values of a$_p$ and a$_n$ are constrained in the a$_p$ - a$_n$ plane to the inside of a 
band defined by two parallel lines of slope $<S_n>$/$<S_p>$ = -0.247. In order to compare with other experiments a WIMP mass of M$_W$  =  50 GeV/c$^2$  was chosen as a reference and the results are shown in  Fig. \ref{Fig:7}. 
\section{Conclusions}\label{sec:conclusions}
	 Significant progress has been achieved recently in PICASSO. 
The analysis of  a first set of  two out of 32 new-generation detectors 
resulted in improved limits for spin-dependent neutralino interactions 
of 0.16 pb on protons and 2.60 pb on neutrons at  90\% C.L. for a WIMP mass of 24 GeVc$^{-2}$. In the WIMP-proton spin-dependent sector the new limits rule out a substantial part of the allowed parameter space of the DAMA/LIBRA experiment if no ion channelling is assumed, but leave still some room if the ion channelling hypothesis is adopted \cite{bernabei,savage}.\\ 
	The main differences and improvements with respect to our previous published results are: 1)  an increase in droplet size by a factor of 20 and a factor of 10 increase in active mass per detector; 2) an improved purification technique resulting in a reduced alpha background; 3) a new discrimination of particle induced events from non-particle produced events by 2D cuts on a new set of variables which are related to the signal energy and frequency content; 4) a new discrimination of neutrons from alpha particles. This last feature has not been exploited yet and is the subject of analysis work in progress. With a full analysis using this latter feature and extended to the full set of detectors we expect a substantial further increase in sensitivity.

\section{Acknowledgements}
We wish to acknowledge the support of the National Sciences and Engineering Research Council of Canada (NSERC), the Canada Foundation for Innovation (CFI),  the National Science Foundation (NSF) for funding and the Czech Ministry of Education, Youth and Sports within the project MSM6840770029 for financial support. We also thank SNOLAB and its staff for its hospitality and for providing help and competent advice whenever needed. This work is also supported by the NSF grant PHY-0856273 and the Indiana University South Bend Research and Development Committee.

\end{document}